\documentclass[aps,pra,twocolumn,amsmath,amssymb,superscriptaddress,10pt]{revtex4-2}
\usepackage[english]{babel}
\usepackage{latexsym}
\usepackage{graphics}
\usepackage{epsfig}
\usepackage{color}
\usepackage{bm}
\usepackage{amsmath}
\usepackage{amssymb}
\usepackage{physics}
\usepackage{hyperref}
\hypersetup{
   colorlinks = true,
   linkcolor=blue,
   citecolor=blue,
   urlcolor=blue
 }

\begin{document}

\title{Vortex dipoles in expanding shell-shaped Bose-Einstein condensates}

\author{A. Tononi}
\affiliation{Department de F\'isica, Universitat Polit\`ecnica de Catalunya, Campus Nord B4-B5, E-08034, Barcelona, Spain}

\begin{abstract}
Releasing shell-shaped Bose-Einstein condensates from the external confinement produces a spherically symmetric density distribution characterized by concentric ripples surrounding a central peak.
Here we investigate how a vortex-antivortex dipole affects this dynamics, finding that increasing dipole separation progressively breaks the spherical symmetry.
Correspondingly, the interplay of vortex physics and curvature produces a nonmonotonic behavior of the cloud aspect ratio with the vortex distance, and decreases the maximal central density achieved during the expansion.
These features can be used for detecting vortex dipoles prepared in shell-shaped superfluids, as well as {a basis} for analyzing their signatures in other thin superfluids with more general curved geometries.
\end{abstract}

\maketitle

Shell-shaped Bose-Einstein condensates are nowadays routinely realized by confining bosonic gases in hollow shells with magnetic and optical potentials \cite{dubessy2025, carollo2022observation, jia2022, huang2025, veyron2026}. 
Interestingly, these thin shells must have zero net vorticity due to topological constraints on the phase distribution of the superfluid order parameter \cite{Turner2010VorticesOC}.
This fact implies that the simplest vortex configuration allowed is a vortex-antivortex dipole \cite{Padavic2020VortexAntivortex}. 
Such dipoles are the simplest topological excitations of shell-shaped gases, and several studies focused on vortex dynamics in spherical superfluids \cite{Turner2010VorticesOC,
Padavic2020VortexAntivortex,
Bereta2021SuperfluidVortex,
Song2022MachineXY,
kanai2021true,
ruban2021bubbles,
ruban2022systems,
xiong2024hydrodynamics,
white2024triangular,
he2023vortex,
cuadra2026,
Caracanhas2022Ellipsoid, gladilin2008, tempere2009}, as well as in classical fluids
\cite{bogomolov1977dynamics,
kimura1987vortex,
kimura1999vortex,
newton2001,
kim2010latitudinal}, 
and discussed their role in the superfluid-to-normal transition on the sphere \cite{kotsubo1984, kotsubo1986, ovrut1991, tononi2022}.

Experimentally, vorticity is typically imaged by allowing the gas to expand, and by comparing its phase profile with that of a reference condensate \cite{corman2014, delpace2022, ciszak2026cooperative}.
In shell-shaped gases, the necessity of implementing such measurement schemes is reduced by the possibility of observing the peculiar self-interference pattern during the shell expansion \cite{lannert2007dynamics}.
While this process has been extensively studied in both experiments \cite{carollo2022observation, jia2022} and theory \cite{lannert2007dynamics, tononi2020, boegel2023controlled, tononi2025}, and despite numerous investigations of vortex dynamics in spherical superfluids \cite{Turner2010VorticesOC,
Padavic2020VortexAntivortex,
Bereta2021SuperfluidVortex,
Song2022MachineXY,
kanai2021true,
ruban2021bubbles,
ruban2022systems,
xiong2024hydrodynamics,
white2024triangular,
he2023vortex,
cuadra2026,
Caracanhas2022Ellipsoid, gladilin2008, tempere2009}, the signatures of vortices in expanding shell-shaped condensates remain unexplored.

\begin{figure}[hbtp]
\centering
\includegraphics[width=0.54\columnwidth]{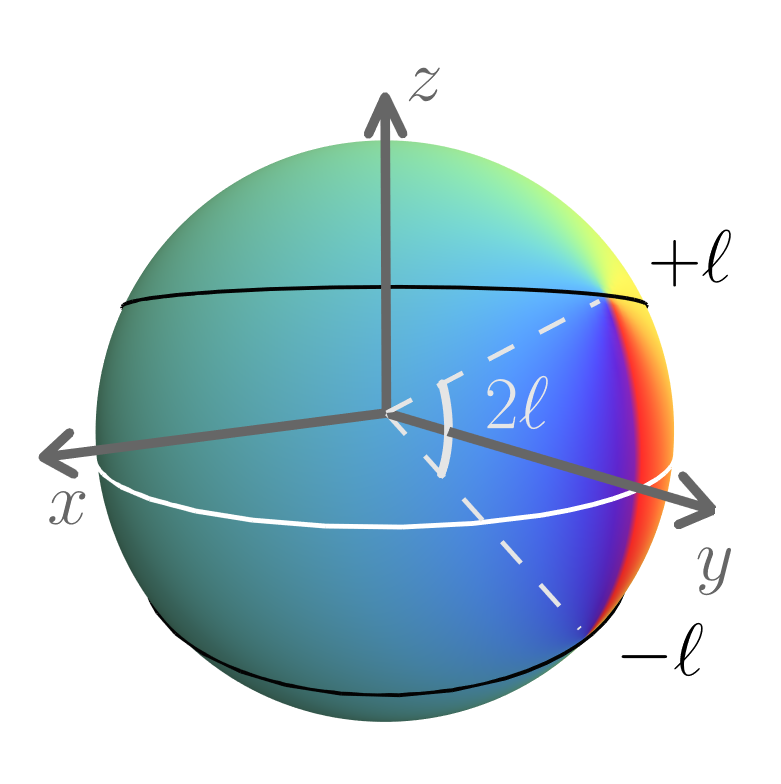}
\caption{System sketch, representing the phase field of a thin spherical superfluid shell hosting a vortex-antivortex dipole confined to the $yz$ plane. 
The vortices are located at latitudes $\pm \ell$ and move along circular trajectories (black circles) parallel to the equator (white circle), maintaining a fixed angular separation $2\ell$.
Upon release from the trap, the condensate expands freely in three dimensions, generating a self-interference pattern which depends on the dipole separation $2\ell \in [0,\pi]$.}
\label{fig1}
\end{figure}

Here we analyze the free expansion of thin spherical shells with a phase profile describing two point vortices with opposite circulation.
Once shell geometry and other basic gas parameters are fixed, the only relevant scale for describing the system is the vortex-antivortex angular separation $2\ell$, defined as the angular distance between a vortex at latitude $\ell$ and an antivortex at latitude $-\ell$ (see Fig.~\ref{fig1}). 
We find that, during the gas expansion, vortex cores produce density holes around the latitudes $\pm \ell$, locally broadening the cloud in the direction perpendicular to the surface.
Even though this phenomenology is compatible with previous studies on flat low-dimensional superfluids \cite{lundh1998, dalfovo2000, madison2000, hosten2003, teles2013}, the observed interference patterns significantly differ, because varying $2\ell$ displaces vortices on a curved surface.
In particular, while vortices located near the poles enhance the density spread in the equatorial plane, vortices near the equator broaden the cloud more along the vertical direction.
This interplay of vortex physics and geometry leads to a nonmonotonic dependence of the cloud aspect ratio on $\ell$, a feature absent in flat low-dimensional systems. 

The cloud aspect ratio can be determined through standard detection techniques such as absorption imaging and the Abel transform for axially symmetric densities \cite{huang2025}.
Combined with measurements of the maximal central density \cite{tononi2025}, this observable enables the detection of vortices prepared under controlled initial conditions, for which various methods are available.
These include phase imprinting of closely spaced vortex dipoles via external potentials \cite{leanhardt2002, murray2013probing}, combining rotation \cite{madison2000} with a pinning laser to displace the vortex dipole along the shell, and the chopstick technique \cite{samson2016}, in which two laser beams are focused on one of the two intersection points on a sufficiently large shell surface \cite{TononiCabrera2026}.

The measurements set provided in this Letter can be extended to evaluate the signatures of vorticity in superfluid films with general curved geometries.
Indeed, given the analytical phase profile of the superfluid \cite{guenther2017, massignan2019, guenther2020, Caracanhas2022Ellipsoid}, it is sufficient to setup the desired configuration of point vortices in the initial condition, and to compute the time evolution through the Gross-Pitaevskii equation.

\vspace{-2mm}
\begin{figure*}[hbtp]
\centering
\includegraphics[width=2.01\columnwidth]{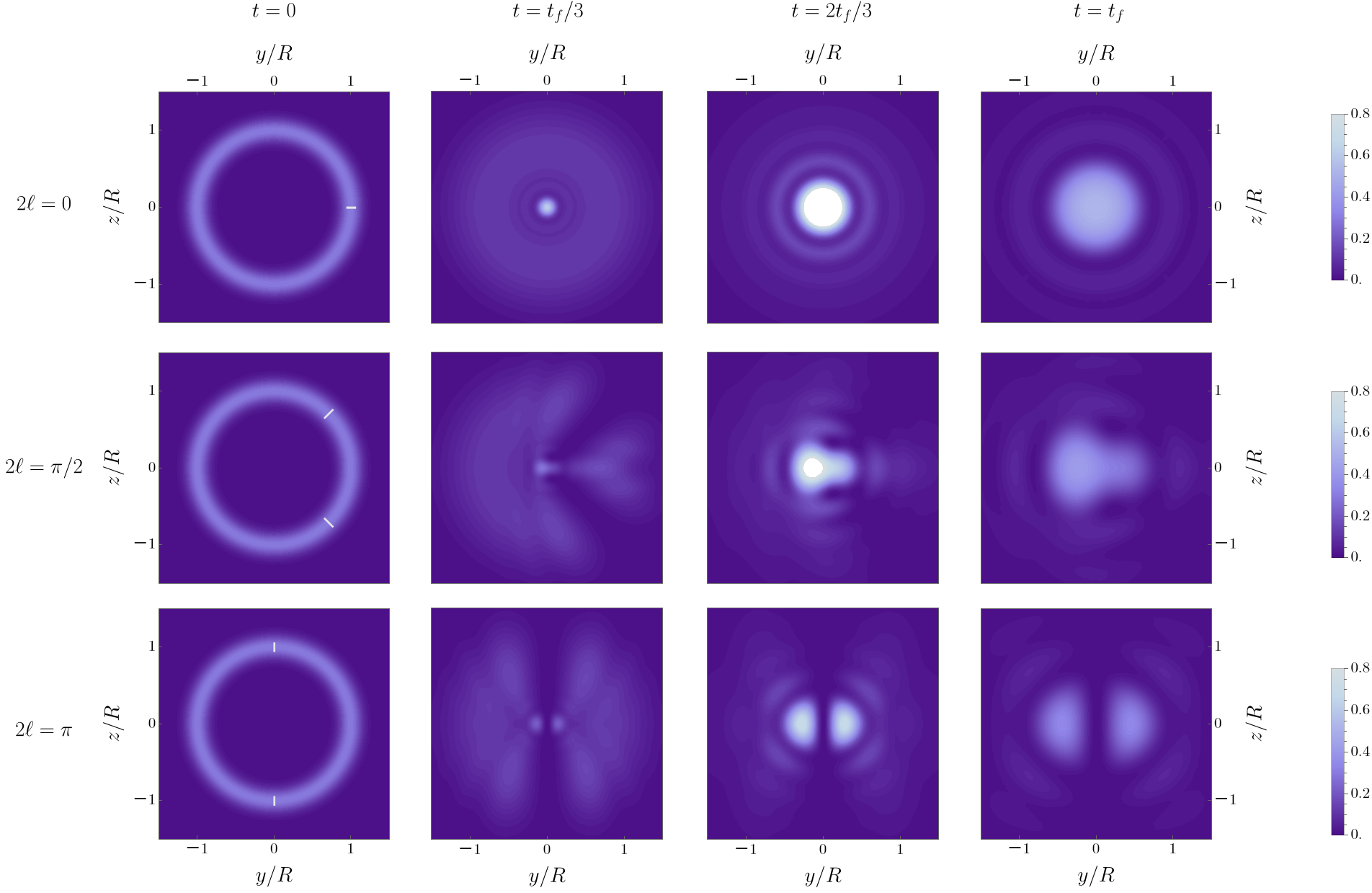}
\caption{Cuts of the normalized density $n(\mathbf{r},t)$ in the $yz$ plane during the free expansion of a shell-shaped Bose-Einstein condensate hosting a vortex-antivortex dipole.
Columns correspond to increasing simulation times $t = \{0, t_f/3, 2 t_f/3, t_f  \}$, with $t_f$ the final time, rows correspond to increasing angular separation of the dipole $2\ell = \{ 0, \pi/2, \pi \}$, and the initial vortex positions are indicated by white marks (cf.~Fig.~\ref{fig1}). 
For coincident vortices ($2\ell = 0$), annihilation leads to a spherically symmetric density distribution. 
Instead, for finite separations ($0 < 2\ell \leq \pi$), the expanding vortex cores produce density holes and result in asymmetric expansion patterns.
We simulate here  Eq.~\eqref{initialstate} dynamics under Eq.~\eqref{3DGPE}, using the parameters: $l_r/R = 0.15$, $N=10^4$, $a/R = 3 \times 10^{-4}$, and $t_{\mathrm{f}} = 0.18 \, t_R$. 
These values are compatible with current experimental realizations of shell-shaped condensates \cite{jia2022, huang2025}, as can be verified by restoring physical units through the unit length $R=10^{-5}\,\mathrm{m}$ and the unit $^{23}\text{Na}$ mass.
}
\label{fig2}
\end{figure*}

\textit{Shell expansion with a vortex dipole.}
We consider a Bose-Einstein condensate initially confined in a shell-shaped configuration.
We model the confinement as a radially shifted harmonic trap centered around the radius $R$, corresponding to the external potential $U(r) = m \omega_{r}^2 (r-R)^2/2$, where $m$ is the atomic mass and $\omega_r$ the radial confinement frequency.
In particular, we consider a radius $R$ sufficiently larger than the oscillator length $l_r = [\hbar/(m \omega_r)]^{1/2}$, which sets the characteristic thickness of the shell.
In this regime, a weakly interacting condensate is well described by a macroscopic wave function in the factorized form \cite{tononi2025}
\begin{equation}
\Psi(r,\theta,\varphi, t=0) = \phi_0(r) \, \psi(\Omega),
\label{initialstate}
\end{equation}
where $\phi_0(r)= \pi^{-1/4} \, l_r^{-1/2} \, e^{-(r-R)^2/(2l_r^2)}$ is the radial ground-state wave function, and $\psi(\Omega)$ is an angular wave function, with $\Omega$ the solid angle. 
Specifically, we parametrize the spherical surface through the standard spherical coordinates $\Omega = (\theta,\varphi) \in [0,\pi] \times [0,2\pi]$.
The approximate form of Eq.~\eqref{initialstate} holds when the effective energy of interactions along the spherical surface is small compared to the radial confinement energy, a condition implying $N a l_r / (\sqrt{2\pi} R^2) \ll 1$ \cite{tononi2025}, where $a$ is the $s$-wave scattering length. 
The simulation parameters chosen in this work satisfy this criterion.

We aim to analyze how specific angular phase profiles of the shell affect the dynamics of the initial state \eqref{initialstate}. 
For this scope, we further decompose the angular wave function $\psi(\Omega)$ as $\psi(\Omega) = \sqrt{n} \, e^{i \Phi(\theta,\varphi)}$, where $n = N/(4\pi R^2)$ is the uniform surface density of $N$ atoms on the sphere, and $\Phi(\theta,\varphi)$ is the phase field. 
In particular, we consider here the phase profile generated by a vortex-antivortex dipole on the spherical shell.

In a flat two-dimensional plane spanned by the complex variable $w$, the phase field is expressed as $\Phi(w) = \Im[F(w)]$, with $F(w) = \ln(w-w_+) - \ln(w-w_-)$ the complex potential of a vortex-antivortex dipole with cores at $w_{\pm}$ \cite{Bereta2021SuperfluidVortex}.
Then, the phase field of the spherical superfluid is obtained by mapping the plane result to the sphere via the stereographic projection $w = 2R \tan(\theta/2) e^{i\varphi}$ and by using the relation $\text{Im}[\ln(w)] = \arg(w)$, which yields \cite{Bereta2021SuperfluidVortex}
\begin{equation}
\Phi(\theta,\varphi) = \arg \left[ \frac{\tan(\theta/2) e^{i\varphi} - \tan(\theta_+/2)e^{i\varphi_+}}{\tan(\theta/2) e^{i\varphi} - \tan(\theta_-/2)e^{i\varphi_-}}  \right],
\label{phase}
\end{equation}
where we have set $w_{\pm} = 2R \tan(\theta_{\pm}/2) e^{i\varphi_{\pm}}$, and $(\theta_{\pm},\varphi_{\pm})$ denote the positions of the vortex cores in spherical coordinates. 
As illustrated in Fig.~\ref{fig1}, we exploit the spherical symmetry by placing the vortices in the $yz$ plane, i.e.~we choose $\varphi_{\pm} = \pi/2$, and we locate them symmetrically with respect to the equator by setting $\theta_{\pm} = \pi/2 \pm \ell$, where $\ell$ is the latitude measured from the equatorial plane.

If the trapping potential $U(r)$ were maintained on, the vortex dipole represented in Fig.~\ref{fig1} would then move parallel to the equator along the black circles.
Namely, each vortex would move in a counterclockwise direction (when viewed from the north pole) and preserve its latitude $\ell$.
Here we instead suppose that the external confinement is suddenly removed, i.e.~we set $U(r) \to 0$ at $t=0$, and we analyze the subsequent free expansion of the condensate.

For this scope, we numerically solve the three-dimensional Gross-Pitaevskii equation
\begin{equation}
i\hbar\dot\Psi(\mathbf{r},t) = \left [ -\frac{\hbar^2\nabla^2}{2m} +g |\Psi(\mathbf{r},t)|^2  \right ]\Psi(\mathbf{r},t),
\label{3DGPE}
\end{equation}
where $g = 4\pi \hbar^2 a /m$ is the coupling constant.
In particular, we analyze the expansion of the state \eqref{initialstate}, with initial phase given by Eq.~\eqref{phase}, by simulating Eq.~\eqref{3DGPE} with standard split-step finite-difference methods that evaluate the Laplacian through the fast Fourier transform.
Simulations use discretized three-dimensional boxes of side $6R$ with up to 250 points in each Cartesian direction, which guarantees a sufficiently dense sampling of the initial shell distribution, and allows checking for results convergence by downsizing the grid.
The following results are expressed in rescaled units, namely lengths are expressed in units of $R$, energies in units of $E_R = \hbar^2/(mR^2)$, and times in units of $t_R = mR^2/\hbar$.

Figure \ref{fig2} illustrates the expansion dynamics of a Bose-Einstein condensate shell with a vortex-antivortex dipole initially located at the angular distance $2\ell$.
Specifically, we plot sections of the normalized density $n(\mathbf{r},t) = |\Psi(\mathbf{r},t)|^2/(N/R^2)$ in the $yz$ plane, considering the dipole angular separations $2\ell = \{0, \pi/2, \pi\}$, and indicating with white marks the initial vortex-core positions.
Note that, for $2\ell = 0$, the vortices coincide and effectively annihilate, yielding a uniform phase field $\Phi = 0$ for all solid angles.
In this case, therefore, the expansion is spherically symmetric and coincident with the one of vortex-free shells, characterized by a central density peak surrounded by concentric interference fringes, whose spacing is set by the interaction strength \cite{lannert2007dynamics, tononi2020}.
For finite separations $0 < 2\ell < \pi$, we observe the rapid expansion of vortex cores, which generate density holes in the cloud \cite{lundh1998, dalfovo2000, madison2000, hosten2003, teles2013}.
This process breaks the spherical symmetry and the density distribution significantly differs from the vortex-free expansion pattern.
Finally, when $2\ell = \pi$, the dipole configuration is symmetric with respect to rotations around the $z$ axis, and the resulting density distribution is also axially symmetric.
We also obtain the density cuts in the $xz$ plane, reported in Fig.~S1 of Supplemental Material (SM) \cite{sm}, finding results consistent with this qualitative analysis.
In the next section, we provide a few quantities to better characterize how vortices affect the shell expansion.

\begin{figure}[hbtp]
\centering
\includegraphics[width=0.99\columnwidth]{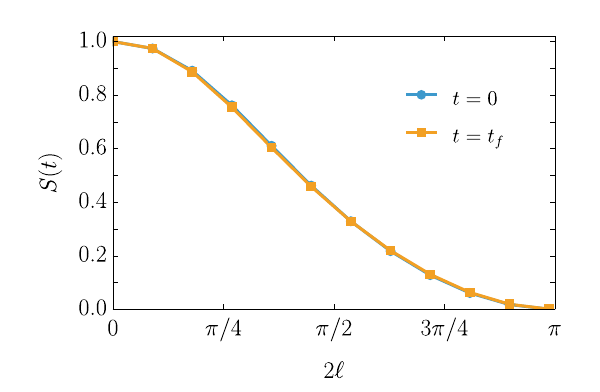}
\caption{Occupation $S(t)$ of the spherically symmetric component vs the vortex-antivortex angular distance $2\ell$, reported at the initial and at the final simulation time. 
Note that increasing the vortex latitude reduces the cloud sphericity, and that nonlinear interactions of the Gross-Pitaevskii equation do not significantly transfer population from excited angular momentum states into the condensate during the dynamics (see the main text).
We choose here simulation parameters as in Fig.~\ref{fig2}.
}
\label{fig3}
\end{figure}

\textit{Sphericity, aspect ratio, and maximal central density. }
The loss of the spherical symmetry with increasing $\ell$ can be quantitatively {characterized in the simulated data} by evaluating the wave function sphericity. 
For this scope, it is sufficient to calculate the fractional occupation of the spherically symmetric angular state $Y_{00}(\Omega)$, where $Y_{lm_l}(\Omega)$ are the spherical harmonics.
In particular, we define the sphericity $S$ as 
\begin{equation}
S(t) = \frac{1}{N} \int_{0}^{\infty} dr \, r^2 |\chi_{00}(r,t)|^2,
\end{equation}
where $\chi_{00}(r,t) = \int d\Omega \, Y_{00}^{*}(\Omega) \Psi(\mathbf{r},t)$ is the radial wave function component with spherical symmetry.
We plot $S(t)$ versus the dipole angular size $2\ell$ in Fig.~\ref{fig3} at the initial time $t=0$, and at the final simulation time $t=t_f$.

We first note that the shell sphericity rapidly reduces as vortices are displaced at larger angular distances.
Moreover, our simulations show that the initial $S(0)$ is basically coincident with $S(t_f)$, i.e.~we find that the sphericity remains constant during the dynamics. 
Indeed, the transfer of angular momentum modes $Y_{l>0 m_l}(\Omega)$ into the spherically symmetric mode $Y_{00}(\Omega)$ occurs only through to the nonlinear term of the Gross-Pitaevskii equation.
However, since this term acts mostly in regions of large density, the conversion is not occurring for the considered potential.
Indeed, the shell density quickly drops as soon as the trap is removed at $t=0$, and then it grows to a large value in the region around the origin.
While the spherically symmetric component $\chi_{00}$ reaches the origin, the finite-$l$ wave-function components face a repulsive centrifugal barrier $V_l(r) \propto l(l+1)/r^2$ during their expansion towards the origin \cite{tononi2025}, which causes them to reflect at a finite radius $r_l/R \propto \sqrt{l(l+1)}$. 
Therefore, these components do not actually reach the origin, and effectively decouple from the spherically symmetric mode.
As a result, the sphericity $S(t)$ remains constant, and it thus provides a good indication of the initial vortex dipole configuration, at least when the gas interactions are sufficiently weak.
However, although $S(t)$ illustrates clearly the results of simulations, experimental accessibility on the wave function is usually limited.

\begin{figure}[t]
\centering
\includegraphics[width=0.99\columnwidth]{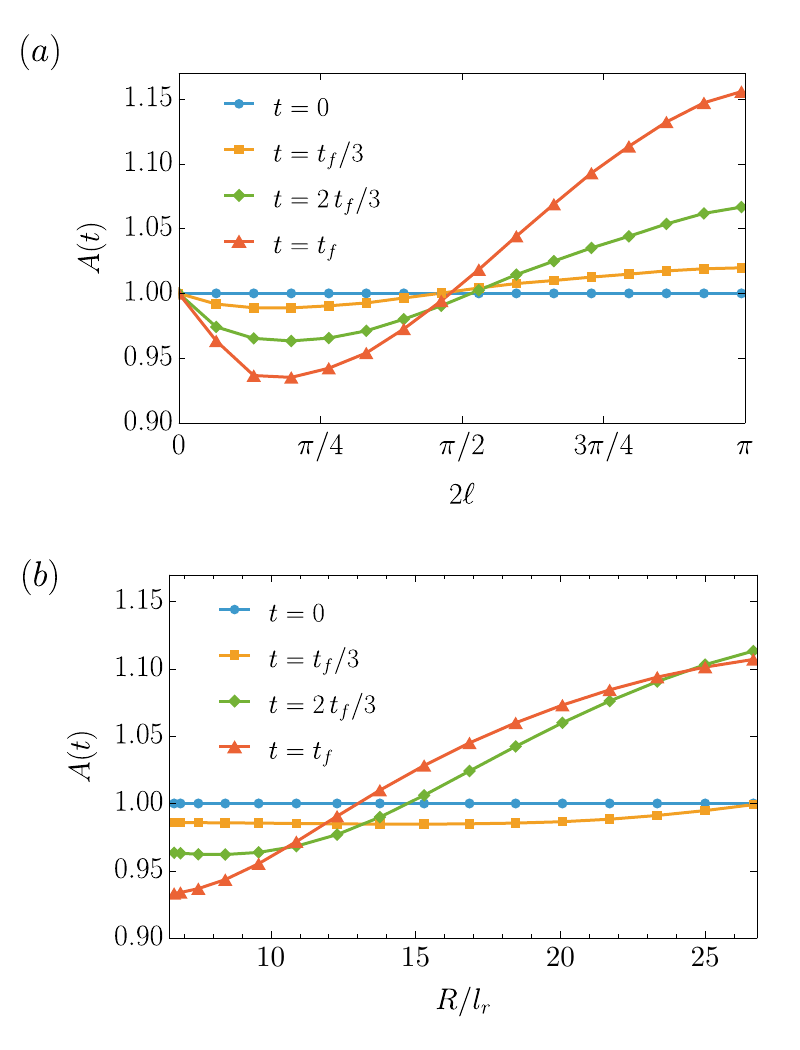}
\caption{
Cloud aspect ratio $A(t)$ as a function of: (a) the dipole angular distance $2\ell$ and for $l_r/R = 0.15$, and (b) the ratio $R/l_r$ and for $2\ell = 0.6$ [corresponding to the minimum of $A(t)$ in (a)].
All other simulations parameters not specified here are chosen as in Figs.~\ref{fig2} and \ref{fig3}.
(a). The aspect ratio shows a nonmonotonic behavior since vortices closer than $2\ell \sim \pi/2$ expand mostly along $x$ and $z$, while vortices further than $2\ell \sim \pi/2$ mostly expand along $x$ and $y$.
(b). The nonmonotonic behavior is reduced when considering thinner shells, that also expand more rapidly.
}
\label{fig4}
\end{figure}

Therefore, let us also characterize the shell expansion through two observables that depend only on the gas density.
For this scope, we first analyze the aspect ratio of the expanding shell.
It is particularly useful in the current geometry to evaluate it as the ratio
\begin{equation}
A(t) = \langle y^2 \rangle_t/\langle z^2\rangle_t,
\label{aspectratio}
\end{equation}
where the density averages are defined as $\langle f(\mathbf{r}) \rangle_t = \int d\mathbf{r} \, f(\mathbf{r}) n(\mathbf{r},t)$. 
We plot in Fig.~\ref{fig4}(a) the aspect ratio $A$ as a function of the vortex-antivortex distance $2\ell$ for different expansion times.
Its non monotonic behavior can be easily understood through geometric considerations.
Indeed, cores of close-by vortices ($2\ell \ll \pi$) mostly push the cloud expansion along the $x$ and $z$ directions (cf. Fig.~\ref{fig1}), leading to a decrease in time of $A$.
In contrast, vortex cores near the poles ($2\ell \sim \pi$) mostly push the cloud expansion along $x$ and $y$, so that the ratio $A$ increases in time. 
For sufficiently weak interactions and small atom numbers, the nonlinear terms are weak and the change in the behavior of $A$ reasonably occur at the value $2\ell \sim \pi/2$, at which vortices are located half-way between the poles and the equator.

However, when reducing the shell thickness and therefore increasing the initial kinetic energy of the shell, the magnitude of the non-monotonic behavior is diminished.
This effect can be appreciated in Fig.~\ref{fig4}(b) which shows $A$ versus the ratio between the radius and the shell thickness $R/l_r$.
Indeed, while for $2\ell = 0.6$ a nonmonotonic region persists for a short time for thicker shells, a more complex dynamics of the cloud aspect ratio, driven by the stronger role of nonlinearity and by the larger initial kinetic energy, reduces this region. 
Analyses based on the aspect ratio $A$, and similar observable ratios, can thus help experiments to reconstruct vortex location in thicker shells, while quantitative side-by-side comparison with the simulated shell sizes is needed to image vortices in the deep two-dimensional regime.

We emphasize that the aspect-ratio variations displayed in Fig.~\ref{fig4} can be realistically observed in current experiments. 
Indeed, time of flight measurements of shell-shaped Bose-Einstein condensates could resolve cloud-size variations at the few-percent level \cite{huang2025}, and longer expansion times $t_f$ would yield even larger signals than the $15\%$ variation displayed in Fig.~\ref{fig4}. 
Furthermore, we verify that moderate density inhomogeneities do not qualitatively modify the behavior reported in Fig.~\ref{fig4} (cf. SM Fig.~S2 \cite{sm}), indicating that the aspect ratio is a robust observable against deviations of the initial density distribution from perfect sphericity. 
Similarly, we verified that evaluating the aspect ratio $A(t)$ by using $x$-column densities does not alter the qualitative behavior displayed in Fig.~\ref{fig4}. 
Finally, our analysis assumes that vortex configurations are placed under controlled experimental conditions as in Fig.~\ref{fig1}, and that the imaging axis aligns with the $x$ direction.

The second detection method based on cloud density measurements consists in extracting the maximal central density achieved inside the angular region $r/R<0.3$ during the expansion \footnote{Note that our considerations are qualitatively unchanged for different choices of the boundary region $r/R = 0.3$.}. 
Consistently with our previous analyses \cite{tononi2025}, we find that this quantity shows a clear decreasing behavior with $2\ell$ for thicker shells corresponding to the ratio $l_r/R \sim 0.15$.
This result can be interpreted as the increasing difficulty of the gas to reach the center when a larger fraction occupies highly-excited angular momentum states $l$, which face the filtering by the repulsive centrifugal potential $V_l (r)$.
At the same time, the dependence of the maximal central density with $2\ell$ becomes nonmonotonic, and therefore less informative on the underlying physical processes when considering thinner shells having $l_r/R \ll 1$.
Overall, this analysis suggests that the maximal central density can be used for inferring the vortex-antivortex distance for thicker and weakly interacting shells as those of Fig.~\ref{fig2}, while it is less reliable for inferring the vortex core positions otherwise.

\textit{Conclusions. }
In this Letter, we have studied the expansion dynamics of a shell-shaped Bose-Einstein condensate whose initial phase profile contains a vortex-antivortex dipole at the angular separation $2\ell$.
As a main result, we find that the expanding dipole in such curved geometry reduces the cloud sphericity when increasing $2\ell$.
Correspondingly, we observe a nonmonotonic behavior of the shell aspect ratio, a fact which is more evident in thicker and weakly interacting shells, and which is explained by the interplay between vortex physics and geometry.

These results can be observed experimentally by studying the expansion of vortex-antivortex dipoles pinned at a controllable distance in shell-shaped Bose-Einstein condensates (see Fig.~\ref{fig1}), and imaging the cloud expansion along an orthogonal axis to the vortex plane.
Possible experimental platforms include binary Bose mixtures in the phase-separation regime \cite{jia2022, huang2025, derrico2019, cavicchioli2025} (see also Refs.~\cite{wolf2022, ma2025, ancilotto2025, ma2025b, veyron2026}), for which we tailor our numerical parameters.

Our analysis contributes to a better understanding of vortex dynamics in curved geometries, which is one of the main motivations for studying shell-shaped gases \cite{carollo2022observation}.
In particular, the current framework can be adapted to simulate multivortex configurations, and to model the free expansion of superfluids with known vortical phase distribution in different geometries, including the cylinder \cite{guenther2017}, the cone \cite{massignan2019}, the torus and surfaces of revolution \cite{guenther2020}, and the ellipsoid \cite{Caracanhas2022Ellipsoid}.

In all these cases, one expects qualitatively similar behavior: Expanding vortex cores generate density holes \cite{lundh1998, dalfovo2000, madison2000, hosten2003, teles2013}, leading to different aspect ratios from the vortex-free case, with geometry-dependent anisotropies determined by the surface curvature.
For instance, in an expanding ellipsoidal shell, a vortex dipole perpendicular to the equator is expected to reduce the aspect ratio when vortices lie near the equator, and to enhance it when they are close to the poles.

\textit{Acknowledgments.}
I thank Gregory E. Astrakharchik and Pietro Massignan for a careful reading of the manuscript, acknowledge discussions with Cesar Cabrera and Andrea Richaud, and within the COST Action SCALES (CA24139), supported by COST (European Cooperation in Science and Technology). I also acknowledge funding by the European Union under the Horizon Europe MSCA programme via Project No. 101146753 (QUANTIFLAC), and support by the Spanish Ministerio de Ciencia, Innovación y Universidades (Grant No. PID2023-147469NB-C21, financed by MICIU/AEI/10.13039/501100011033 and FEDER-EU).

\textit{Data availability.} The data that support the findings of this
article are openly available \cite{zenodo}.

\bibliographystyle{apsrev4-2}
\bibliography{bibliography.bib}

\setcounter{equation}{0}
\setcounter{figure}{0}
\setcounter{table}{0}

\makeatletter
\renewcommand{\theequation}{S\arabic{equation}}
\renewcommand{\thefigure}{S\arabic{figure}}
\renewcommand{\thetable}{S\arabic{table}}
\makeatother

\newpage
\widetext
\textbf{\center{Supplemental Material for: ``Vortex dipoles in expanding shell-shaped Bose-Einstein condensates''}}

\begin{figure*}[hbtp]
\centering
\includegraphics[width=1.01\columnwidth]{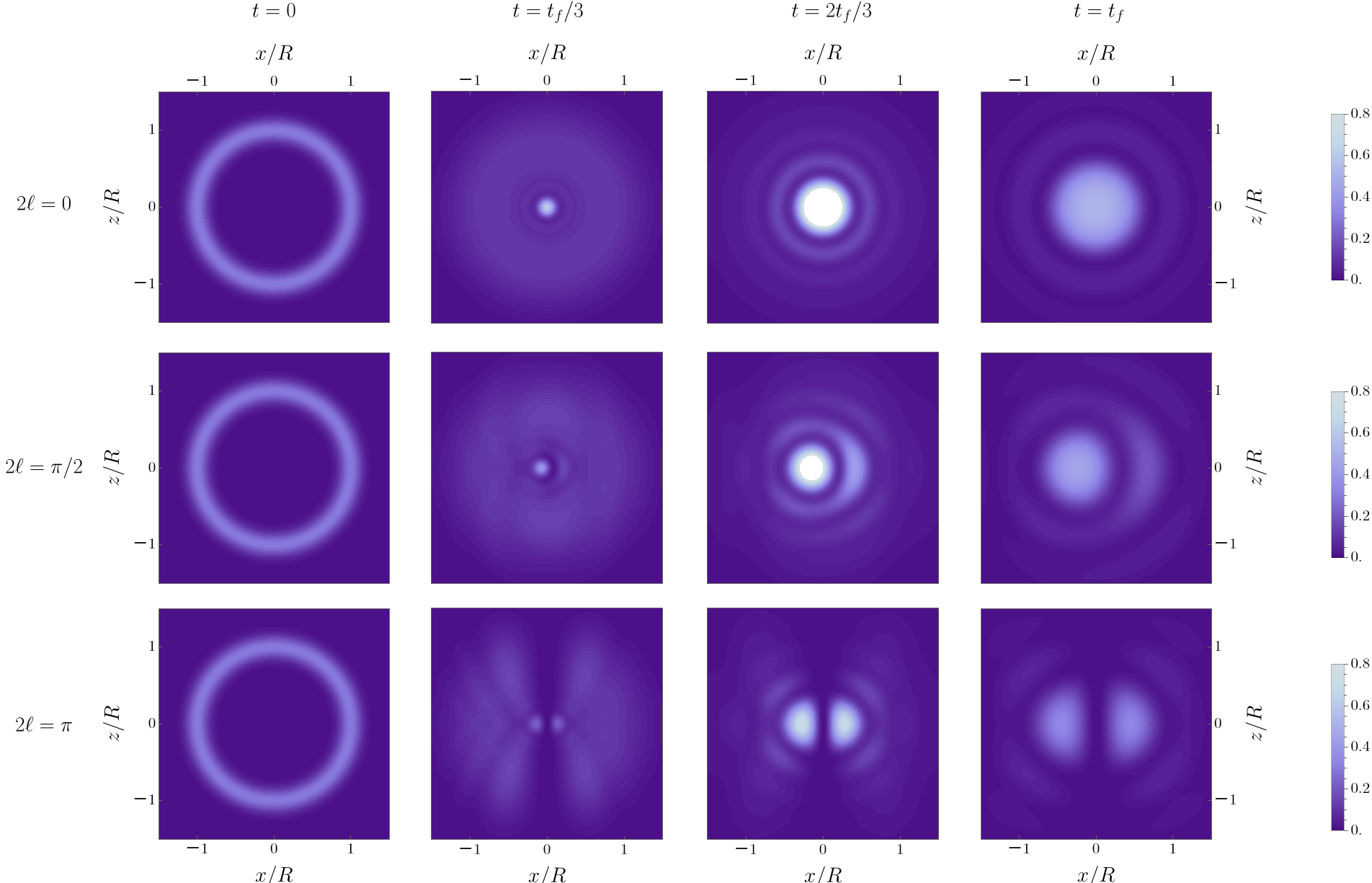}
\caption{Density cuts in the $xz$ plane during the free expansion of a Bose-Einstein condensate shell with vortices at angular distance $2\ell$. We use the same parameters as in Fig.~2.
}
\label{fig5}
\end{figure*}

\vspace{4mm}
To complement Fig.~2 of the main text, we report in Fig.~\ref{fig5} the sections of the normalized density $n(\mathbf{r},t)$ in the $xz$ plane. 
Note how, due to the axial symmetry along $z$, the $xz$ and $yz$ density cuts coincide for $2\ell = 0$ and $2\ell = \pi$.
Instead, the expansion differs markedly for $2\ell = \pi/2$, for which there is no axial symmetry.
Furthermore, the expanding superfluid for $2\ell = \pi/2$ is not symmetric under $x \to -x$.
This effect is a consequence of vortex motion in a counterclockwise direction around the $z$ axis for $0 < 2\ell < \pi$. 
If vortex signs are exchanged, motion occurs instead clockwise, and the reflected $xz$ section is obtained.

\begin{figure}[hbtp]
\centering
\includegraphics[width=0.51\columnwidth]{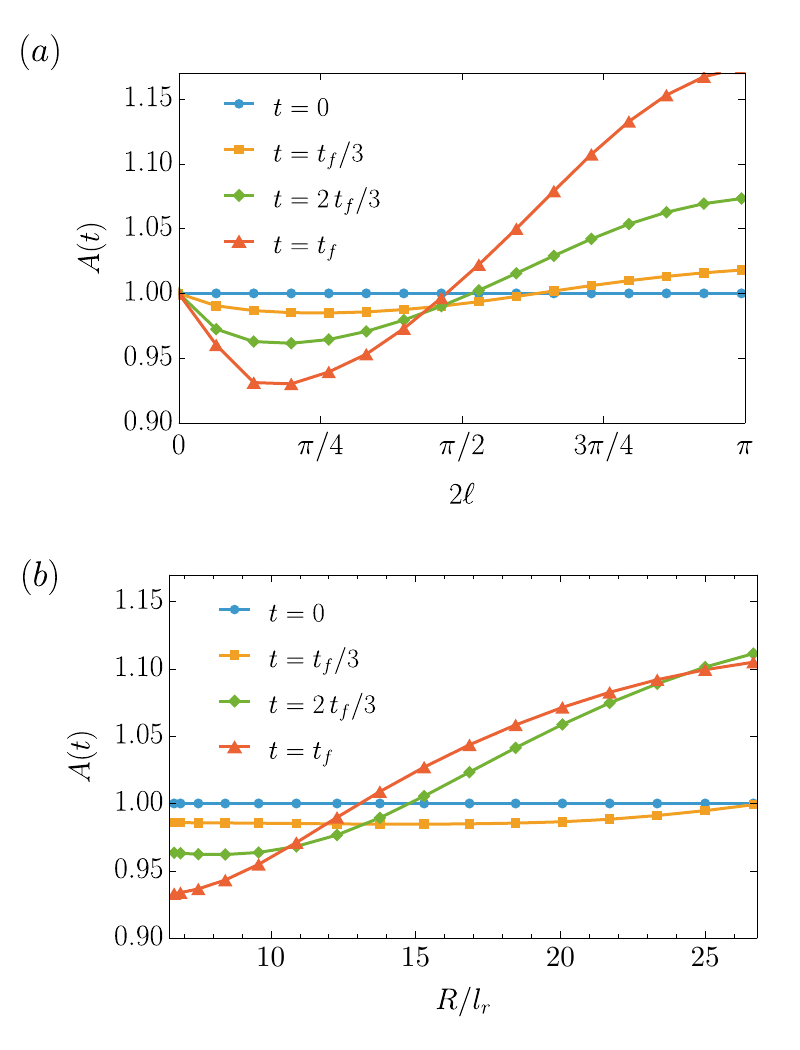}
\caption{
Cloud aspect ratio $A(t)$, corresponding to Fig.~4. All parameters are kept unchanged, except for the inclusion of the $z$-dependent density gradient in the initial state (see the Appendix text). The resulting aspect ratio dynamics is practically the same of Fig.~4.
}
\label{fig6}
\end{figure}

Also, to complement Fig.~4 of the main text, we report in Fig.~\ref{fig6} results on the cloud aspect ratio $A(t)$ of moderately inhomogeneous shells. 
In particular, these simulations use the initial wave function $\Psi(\mathbf{r},0)(1 - 0.1\, z/R)^{1/2}$, with $\Psi(\mathbf{r},0)$ the initial states used in Fig.~4, and all other parameters are kept unchanged.
Note that the aspect ratio dynamics is qualitatively unaffected by this density gradient, which has comparable magnitude to the one of experimental realizations \cite{jia2022}.
These results indicate that the main vortex signatures discussed in the paper are robust against moderate asymmetries of the density distribution, and support the use of the aspect ratio as an observable for characterizing vortex configurations.

\end{document}